\documentclass[aps,prl,twocolumn,groupedaddress,showpacs]{revtex4}
\usepackage{graphicx}
\usepackage{dcolumn}
\usepackage{bm}

\begin{document}


\title{\boldmath 
Interference Study of the $\chi_{c0}(1^3P_0)$ 
in the Reaction $\bar{p}p\rightarrow\pi^0\pi^0$
}


\author{
M.~Andreotti,$^2$
S.~Bagnasco,$^{3,7}$ 
W.~Baldini,$^2$  
D.~Bettoni,$^2$  
G.~Borreani,$^7$  
A.~Buzzo,$^3$ 
R.~Calabrese,$^2$  
R.~Cester,$^7$  
G.~Cibinetto,$^2$
P.~Dalpiaz,$^2$ 
G.~Garzoglio,$^1$
K.~E.~Gollwitzer,$^1$ 
M.~Graham,$^5$
M.~Hu,$^1$
D.~Joffe,$^6$,
J.~Kasper,$^6$  
G.~Lasio,$^{7,4}$  
M.~Lo Vetere,$^3$  
E.~Luppi,$^2$  
M.~Macr\`\i,$^3$  
M.~Mandelkern,$^4$  
F.~Marchetto,$^7$  
M.~Marinelli,$^3$ 
E.~Menichetti,$^7$ 
Z.~Metreveli,$^6$
R.~Mussa,$^{2,7}$  
M.~Negrini,$^2$
M.~M.~Obertino,$^{7,5}$
M.~Pallavicini,$^3$  
N.~Pastrone,$^7$  
C.~Patrignani,$^3$  
T.~K.~Pedlar,$^6$  
S.~Pordes,$^1$
E.~Robutti,$^3$  
W. Roethel,$^{6,4}$
J.~L.~Rosen,$^6$  
P.~Rumerio,$^{7,6}$ 
R.~Rusack,$^{5}$
A.~Santroni,$^3$ 
J.~Schultz,$^4$ 
S.~H.~Seo,$^5$
K.~K.~Seth,$^6$ 
G.~Stancari,$^{1,2}$ 
M.~Stancari,$^{4,2}$
A.~Tomaradze,$^6$
I.~Uman,$^6$
T.~Vidnovic~III,$^{5}$  
S.~Werkema$^1$
and P.~Zweber$^6$\\
\vspace{0.06in}
(Fermilab E835 Collaboration)  \\
\vspace{0.06in}
$^1$Fermi National Accelerator Laboratory, Batavia, Illinois 60510 \\
$^2$Istituto Nazionale di Fisica Nucleare and University of Ferrara, 
44100 Ferrara, Italy \\
$^3$Istituto Nazionale di Fisica Nucleare and University of Genova, 
16146 Genova, Italy \\
$^4$University of California at Irvine, California 92697 \\
$^5$University of Minnesota, Minneapolis, Minnesota 55455\\
$^6$Northwestern University, Evanston, Illinois, 60208 \\
$^7$Istituto Nazionale di Fisica Nucleare and University of Torino,
10125, Torino, Italy}
\vspace{0.06 in}


\begin{abstract}
Fermilab experiment E835 has observed $\bar{p}p$ annihilation production 
of the charmonium state $\chi_{c0}$ and its subsequent decay into 
$\pi^0\pi^0$.
Although the resonant amplitude is an order of magnitude smaller than 
that of the non-resonant continuum production of $\pi^0\pi^0$, an enhanced
interference signal is evident.
A partial wave expansion is used to extract physics parameters.
The amplitudes $J=0$ and $2$, of comparable strength, 
dominate the expansion. 
Both are accessed by $L=1$ in the entrance $\bar{p}p$ channel.
The product of the input and output branching fractions is determined to be
$B(\bar{p}p\rightarrow\chi_{c0})\times B(\chi_{c0}\rightarrow\pi^0\pi^0)=
(5.09\pm0.81\pm0.25)\times 10^{-7}$.

\end{abstract}

\pacs{13.25.Gv;13.75.Cs;14.40.Gx}

\maketitle


E835 studies charmonium formed in $\bar{p}p$ annihilation, 
a technique that allows direct access to all charmonium states.
The experiment, located in the Antiproton Accumulator at Fermilab, 
recorded new data in 2000. 
This run included  $33~$pb$^{-1}$ of luminosity collected at 17 energies
(3340$-$3470~MeV) at the $\chi_{c0}$.
Results from the scan of the $\chi_{c0}$ resonance using the decay channel
$\chi_{c0}\rightarrow J/\psi\,\gamma, J/\psi\rightarrow e^+e^-$
have already been published \cite{psigamma}.
Simultaneously recorded neutral events provide the $\pi^0\pi^0$ data 
reported here. 
The energy and luminosity measurements are the same for the 
$J/\psi\,\gamma$ and $\pi^0\pi^0$ channels.

Charmonium resonances are scanned using a tunable, stochastically 
cooled $\bar{p}$-beam that intersects a  
hydrogen gas jet target \cite{chi2_gg}.
At each energy point, the cross section is measured by normalizing 
the number of events which satisfy the event selection criteria for 
the final state under study to the integrated luminosity collected.
Typically, an excitation curve is extracted from the large hadronic background 
by tagging electromagnetic final states. 
Hence, resonance parameters may be determined without relying on the detector
resolution, but only on the knowledge of the $\bar{p}$ beam.
At any setting of the beam momentum, the $\bar{p}p$ center of mass
energy ($E_{cm}$) is known to 0.2~MeV; 
the spread (r.m.s.) in $E_{cm}$ is on average about 0.4~MeV for these data.
This resolution is substantially better than could be provided by detection
equipment alone and is much smaller than the $\chi_{c0}$ width 
of $9.8~$MeV \cite{psigamma}.
However, when a large non-resonant continuum is present
in the final state of interest, 
as in the $\bar{p}p\rightarrow\pi^0\pi^0$ analysis, 
special techniques must be used to determine the resonance parameters.

In the vicinity of the $\chi_{c0}$, 
the differential cross section for the process $\bar{p}p\rightarrow\pi^0\pi^0$ 
is
\begin{equation}\label{eq:nosums}
\frac{d\sigma}{dz}= \Big|\frac{-A_R}{x+i}+Ae^{\imath\delta_A}\Big|^2+\Big|Be^{\imath\delta_B}\Big|^2,
\end{equation}
%
\begin{equation}\label{eq:hel0}
A\,e^{i\delta_{A}}\equiv\sum_{J=0,2,...}^{J_{max}} (2J+1)~C_J(x)~e^{i\delta_J(x)}~P_J(z),
\end{equation}
\begin{equation}\label{eq:hel1}
B\,e^{i\delta_{B}}\equiv\sum_{J=2,4,...}^{J_{max}} \frac{(2J+1)}{\sqrt{J(J+1)}}~C_J^1(x)~e^{i\delta_J^1(x)}~P_J^1(z),
\end{equation}
where $P_J^M(z)$ are Legendre functions,
$x \equiv 2(E_{CM}-M_{\chi_{c0}})/\Gamma_{\chi_{c0}}$ and
$z \equiv \vert \cos{\theta^*}\vert$, with $\theta^*$ defined as the 
$\pi^0\pi^0$ production angle in the center of mass frame 
with respect to the $\bar{p}$ direction. 

The term $-A_R/(x+i)$ is the parameterization of a 
Breit-Wigner resonant amplitude with $\mathrm{spin}=0$. 
The partial wave 
sums in Eqs.~(\ref{eq:hel0}) and (\ref{eq:hel1}) represent the continuum 
contributions 
resulting from the initial states $\vert\lambda_{\bar{p}}-\lambda_p\vert=0$ 
(helicity-0) and $\vert\lambda_{\bar{p}}-\lambda_p\vert=1$ (helicity-1),
respectively. 
These two helicity states 
are orthogonal and do not interfere with one another.  
The $\chi_{c0}$ is only produced in the helicity-0 initial state.

At fixed $z$, the continuum terms
$A\,e^{i\delta_{A}}$ and $B\,e^{i\delta_{B}}$ do not change 
markedly as the energy varies across the resonance.
It is then useful to
rewrite Eq.~(\ref{eq:nosums}):
\begin{equation} \label{eq:crossterm}  
 \frac{d\sigma}{dz}=\frac{A_R^2}{x^2+1}+A^2+\underbrace{2 A_R A\,\frac{\sin{\delta_{A}}-x\cos{\delta_{A}}}{{x^2+1}}}_{\mathrm{interference-term}}+B^2.
\end{equation}
The above expression shows 
how even a small resonant contribution can lead to a detectable 
interference signal,
albeit superimposed on a large continuum.
For example, if the contribution $A_R^2$ from the resonance at the resonance 
peak energy is 1\% of the size of the cross section from the helicity-0 
continuum $A^2$, the contribution from the factor $2A_R A$ will be 20 times 
larger than $A_R^2$. 
The factor $(\sin{\delta_A}-x\cos{\delta_A})/(x^2+1)$ determines 
the shape of the interference pattern seen in the cross section.

Two independent data analyses were performed and provide consistent
results \cite{Paolo,Ted}.
One of them is presented here. 
The initial data sample consists of 4-photon events,
selected by the neutral trigger (efficiency $\sim96.1\%$)
and the central shower calorimeter.
The energy resolution of this detector is
$\sigma_E/E\simeq6\%/\sqrt{E(\mathrm{GeV})}+1.4\%$,
while the polar and azimuthal angular resolution are 
$\sigma_{\theta}\simeq 6$~mrad and 
$\sigma_{\phi}\simeq 11$~mrad, respectively \cite{e760_etac}. 
A 5\% cut on the confidence level of a 4C kinematic fit to the
$\bar{p}p\rightarrow\gamma\gamma\gamma\gamma$
hypothesis ensures four-momentum conservation.
Of the three possible ways to pair the 4 photons, 
one at most is a candidate for the 
$\bar{p}p\rightarrow\pi^0\pi^0\rightarrow\gamma\gamma\,\gamma\gamma$ 
hypothesis due to the small opening angle of the 2-photon  
$\pi^0$ decays.
A selection was made on the 2-photon invariant masses 
($m_{\gamma_1\gamma_2}$ and $m_{\gamma_3\gamma_4}$) 
and the $\pi^0\pi^0$ colinearity in the c.m. frame. 
The total number of $\pi^0\pi^0$ candidate events is $\sim500,000$. 
The product of the geometric acceptance and selection efficiency
slowly increases up to $z\simeq0.5$ 
(the average over this range is $\sim63\%$) where it starts to decrease rapidly.
The instantaneous luminosity varied substantially during the data taking;
the averages within each energy point were from 
$1.7$ to $3\times10^{31}~$s$^{-1}$cm$^{-2}$.
The event pileup was studied and corrected for by means of random triggers
recorded throughout the data taking.
The rate-dependent loss varied among the energy points
from $14\%$ to $23\%$ with an average of $\sim20\%$.

\begin{figure}[htb]
\includegraphics[width=18pc]{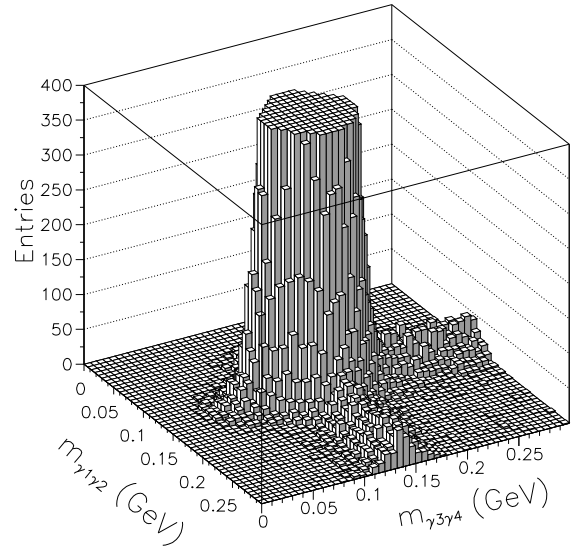}
\caption{\label{fig:2pi0paper_legox40} 
The $m_{\gamma_1\gamma_2}$ versus $m_{\gamma_3\gamma_4}$ LEGO plot in the 
region of the $\pi^0\pi^0$ peak,
which is truncated at about $3\%$ of its height.
}
\end{figure}

The background 
comes mostly from the $\pi^0\pi^0\pi^0$ and 
$\pi^0\omega\rightarrow\pi^0\pi^0\gamma$ channels \cite{e760_2body}, 
and was determined by fitting the LEGO plot in 
Fig.~\ref{fig:2pi0paper_legox40}.
A $\sim$2\% subtraction resulted from this background.

\begin{figure}[htb]
\includegraphics[width=18pc]{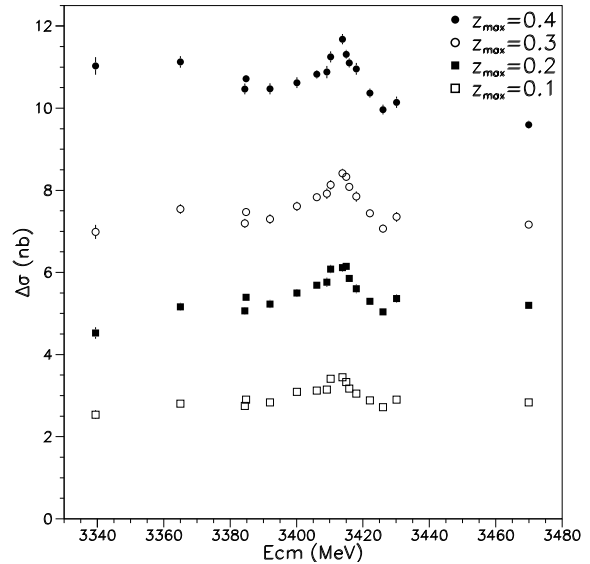}
\caption{\label{fig:2pi0_plotcs_1234} 
The $\bar{p}p\rightarrow\pi^0\pi^0$ cross section 
$\Delta\sigma=\int_0^{z_{max}}(d\sigma/dz)dz$
as a function of $E_{CM}$.
The error bars are statistical.}
\end{figure}

The measured cross section of the $\bar{p}p\rightarrow\pi^0\pi^0$ reaction 
in the $\chi_{c0}$ region as a function of $E_{cm}$
is shown in Fig.~\ref{fig:2pi0_plotcs_1234}. 
A non-resonant $\bar{p} p\rightarrow\pi^0\pi^0$ production with
a smooth energy dependence is present throughout the scanned region.
A clear resonance signal is visible close to the $\chi_{c0}$ mass; 
a finer scale plot shows a peak-shift of $\sim 2$~MeV 
(as a consequence of interference) toward low energy 
with respect to $M_{\chi_{c0}}=3415.4~$MeV/c$^2$ of Ref.~\cite{psigamma}.

\begin{figure}[htb]
\includegraphics[width=20pc,]{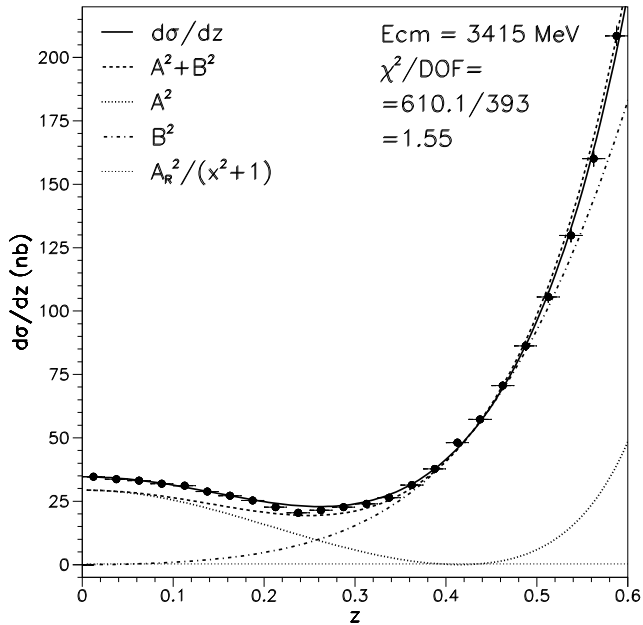}
\caption{\label{fig:2pi0_angdis_fit} The 
$\bar{p}p\rightarrow\pi^0\pi^0$ cross section versus $z$ 
at $E_{CM}=3415~$MeV. 
The fit of Eqs.~(\ref{eq:nosums}), (\ref{eq:hel0}) and (\ref{eq:hel1}), 
which was simultaneously performed on all 17 energy settings,
is shown along with its components.}
\end{figure}

A binned maximum likelihood fit using the parameterization of 
Eqs.~(\ref{eq:nosums}), (\ref{eq:hel0}) and (\ref{eq:hel1}), 
with $J_{max}=4$, was performed simultaneously on 
all energy points.
The $\chi_{c0}$ mass and width are set to the values
(reproduced in Table~\ref{tab:table})
that E835 measured via the reaction 
$\bar{p}p\rightarrow\chi_{c0}\rightarrow J/\psi\,\gamma$, 
$J/\psi \rightarrow e^+e^-$ \cite{psigamma},
which had virtually zero background and non-resonant cross 
section.
The result of the fit is shown 
in Fig.~\ref{fig:2pi0_angdis_fit}. 
The curve $A^2+B^2$ shows the sum of the 
non-resonant cross section contributions.
The effect of the resonance, amplified by the interference, is seen in 
the separation (evident at small $z$) between $d\sigma / dz$ and $A^2+B^2$ 
and is almost entirely due to the interference-term of 
Eq.~(\ref{eq:crossterm}).
The separation decreases as $z$ increases, following the trend 
of $A$. 
The term $B^2$ is small at small values of $z$, 
due to a factor $z$ present in all the associate functions 
$P_J^1(z)$. 
The net suppression factor of $B^2$ with
respect to $A^2$ is $z^2$ at small $z$.
The contribution of 
the ``pure'' resonance, 
$A_R^2/(x^2+1)$, is negligible
and not distinguishable from zero in the figure.

The fit has 408 bins: $17$ energy points times 24 bins in $z$  
(from 0 to 0.6). The number of free parameters is $15$: 
the resonance amplitude, $A_R$; the coefficients $C_{J=0,2,4}$ and 
$C_{J=2,4}^1$ (each of them is given a linear energy dependence); 
and the phases $\delta_{J=0,2,4}$ and $(\delta_4^1-\delta_2^1)$. 
Including energy dependence in the phases does not significantly alter 
nor improve the fit.

The partial wave summation is truncated at $J_{max}=4$.
The addition of a relatively small $J=4$ amplitude 
to a basic $J_{max}=2$ fit 
is already ``fine tuning''.
At $E_{cm}=3415$~MeV the relative amounts are 
$C_0:(C_2,C_2^1):(C_4,C_4^1)\simeq 1:(0.65,0.41):(0.20,0.12)$.
$J=6$ and higher partial waves are highly oscillatory functions and do not
significantly improve the fit nor are physics motivated.
$L_{\bar{p}p}=1$ ``feeds'' $J=0$ and 2.
$L_{\bar{p}p}=3$ is the minimum entrant angular momentum for $J=4$.
Increasing $L_{\bar{p}p}$ correlates directly with the collision impact
parameter ($b$).
The physics of the annihilation process into $\pi^0\pi^0$ at small $z$
strongly favors small $b$ values.
The $\chi_{c0}$ channel obviously requires total valence quark annihilation.
The non-resonant $\pi^0\pi^0$ channel does not require total valence 
quark annihilation; however, hard, short-range collisions are required
to redirect the non-annihilating quarks into reforming as part of the 
$\pi^0\pi^0$ final state.
Our truncating partial wave expansion supports this physical picture.

\begin{figure}[t]
\includegraphics[width=20pc]{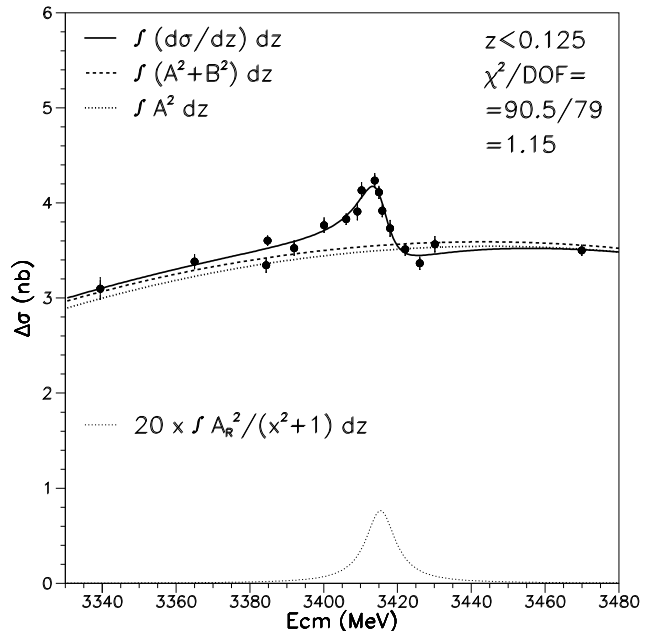}
\caption{\label{fig:2pi0_cs_fit}  
The $\bar{p}p\rightarrow\pi^0\pi^0$ cross section 
$\Delta\sigma=\int_0^{0.125}(d\sigma/dz)dz$
as a function of $E_{CM}$.
The reduced range fit and its components are also shown.}
\end{figure}

In order to extract the product of the branching ratios 
$B(\chi_{c0} \rightarrow \bar{p}p)\times B(\chi_{c0} \rightarrow\pi^0\pi^0)$,
the natural place to focus on is the small $z$ region,
where the $z^2$ suppression of the non-interfering helicity-1
continuum (of magnitude $B^2$, see Eq.~(\ref{eq:nosums})\,) 
with respect to the interfering helicity-0 
continuum (of magnitude $A^2$) is exploited.
At small $z$ the uncertainty on the relative amounts of $A^2$ and $B^2$, 
critical in quantifying the amplification effect of the interference, 
is severely limited and guarantees a model-insensitive result.
In addition, the interference-enhanced $\chi_{c0}$ signal has a substantial 
size at small $z$ and decreases at increasing $z$, while the forward-peaked
continuum production dominates at larger $z$.
Accordingly, a new fit was performed in a reduced range $0<z<0.125$\,.
Eq.~(\ref{eq:nosums}) was used again with the following differences.
First, the small contribution of 
$B^2=\vert Be^{\imath\delta_B}\vert^2$ was fixed and taken from the fit in
Fig.~\ref{fig:2pi0_angdis_fit}, which is dominated by the forward peak
of this component.
Second, a new parameterization of the helicity-0 continuum was introduced;
due to the small range of $z$,
the number of free parameters can be decreased 
by employing a polynomial expansion on $z$ and $x$ 
for $A^2 = a_0+a_1 x +a_2 x^2 +a_3 z^2$.
This is an adequate approximation to the partial wave expansion in the small $z$
interval.
The resulting $d\sigma/dz$, integrated over the reduced range, 
is shown in Fig.~\ref{fig:2pi0_cs_fit}.
As evident from the figure, 
the constant term $a_0$ is dominant, while 
$a_1$ and $a_2$ provide a small slope and curvature as a function of $x$.
The small $z^2$ dependence in the reduced $z$ region is accommodated by $a_3$.
The fit has 6 free parameters 
($A_R$, $a_0$, $a_1$, $a_2$, $a_3$, and $\delta_{A}$) 
for 85 bins  
($17$ energy points times $5$ bins in $z$, from $0$ to $0.125$).
By searching for improvements in the $\chi^2$ it is found that the phase 
$\delta_{A}$ does not exhibit any dependence on $z$ in this small
$z$-range nor on the energy, and that $A^2$ does not require additional
powers of $x$ and $z$, nor mixed terms such as $xz^2$ and $x^2z^2$.

It has been noted that the ``pure'' Breit-Wigner 
(the fictional cross section that would result if the non-resonant amplitudes 
could be turned off) is very small.
In Fig.~\ref{fig:2pi0_cs_fit} it is shown enhanced by 20 to make it
comparable to the signal actually detected.

The fit provides the value of $A_R$, which is related to the product of the
input and output branching ratios by
$A_R^2=\pi \lambdabar^2 \times
B(\chi_{c0} \rightarrow \bar{p}p)\times B(\chi_{c0} \rightarrow\pi^0\pi^0)$,
where $\lambdabar$ is the center of mass de Broglie wavelength 
of the initial state.
%
The result for the product of the input and output branching ratios 
obtained with $M_{\chi_{c0}}$ and $\Gamma_{\chi_{c0}}$ constrained 
to the values obtained in our $J/\psi\,\gamma$ measurement \cite{psigamma}
is reported in Table~\ref{tab:table} as 
``Final result for $B_{in} \times B_{out}$''. 
The dominant systematic error arises from the luminosity determination.
The uncertainty on the knowledge of the helicity-1 continuum affects
$B_{in} \times B_{out}$ by $\sim1\%$.

The difference of phase ($\delta_A$) between the amplitudes of the helicity-0 
continuum and the resonance produces a modest constructive interference on the
low-energy and destructive on the high-energy side of the $\chi_{c0}$ mass.
\begin{table}
\caption{\label{tab:table} E835 Results (errors are statistical and
systematic, respectively). }
\begin{ruledtabular}
\begin{tabular}{c|c|c}
$~~~~~~~~~~~~~B_{in}\equiv$		&\multicolumn{2}{c}{Common channel $B(\chi_{c0}\rightarrow \bar{p}p)$}			\\ 
$~~~~~~~~~~~~B_{out}\equiv$		&$B(\chi_{c0}\rightarrow J/\psi\,\gamma)$\footnotemark[1]	&$B(\chi_{c0}\rightarrow \pi^0\pi^0)$	\\ \hline
$M_{\chi_{c0}}$ (MeV/c$^2$)		&$3415.4\pm0.4\pm0.2$\footnotemark[2]				&$3414.7^{+0.7}_{-0.6}\pm0.2$\footnotemark[3]		\\
$\Gamma_{\chi_{c0}}$ (MeV)		&$9.8\pm1.0\pm0.1$\footnotemark[2]				&$8.6^{+1.7}_{-1.3}\pm0.1$\footnotemark[3]		\\
$B_{in} \times B_{out}$ ($10^{-7}$)	&$27.2\pm1.9\pm1.3$\footnotemark[2]				&$5.42^{+0.91}_{-0.96}\pm0.22$\footnotemark[3]		\\ \hline\hline 
\multicolumn{2}{c|}{Final result for $B_{in} \times B_{out}$ ($10^{-7}$)}				&$5.09\pm0.81\pm0.25$\footnotemark[4]			\\
\multicolumn{2}{c|}{and phase $\delta_A$ (degree)}							&$39\pm5\pm6$\footnotemark[4]				\\  
\end{tabular}
\end{ruledtabular}
\footnotetext[1]{The $J/\psi$ was detected through its decay into $e^+e^-$ \cite{psigamma,psi_e+e-}.}
\footnotetext[2]{From \cite{psigamma}, where $B_{in} \times B_{out}$, $M_{\chi_{c0}}$ and $\Gamma_{\chi_{c0}}$ were free parameters.}
\footnotetext[3]{This analysis with $B_{in} \times B_{out}$, $M_{\chi_{c0}}$ and $\Gamma_{\chi_{c0}}$ as free parameters.}
\footnotetext[4]{This analysis with $M_{\chi_{c0}}$ and $\Gamma_{\chi_{c0}}$ set to values from Ref.~\cite{psigamma}.}
\end{table}

A consistency check is provided by allowing $M_{\chi_{c0}}$ and 
$\Gamma_{\chi_{c0}}$ to vary along with $B_{in} \times B_{out}$ and $\delta_A$.
Table~\ref{tab:table} shows that the resultant $M_{\chi_{c0}}$ and 
$\Gamma_{\chi_{c0}}$ values agree with the $J/\psi\,\gamma$ values. 
Although the $\pi^0\pi^0$ sample is more copious than the $J/\psi\,\gamma$,
the values determined by the $\pi^0\pi^0$ fit have larger uncertainties 
because of the higher number of (coupled) fit parameters.

The results presented so far are obtained from the E835 2000 data sample
alone, with the exception of the well-known 
$B(J/\psi\rightarrow e^+e^-)$ \cite{psi_e+e-}.
To deconstruct the entrance and exit channel branching fractions, we
must use data from the literature.
$B(\chi_{c0}\rightarrow \pi^0\pi^0)$ \cite{chi0_2pi0} is more than an order of
magnitude larger (and thus better measured)
than $B(\chi_{c0}\rightarrow \bar{p}p)$.
We then obtain $B(\chi_{c0}\rightarrow \bar{p}p)=
(2.04\pm0.32_{stat}\pm0.10_{syst}\pm0.28_{PDG})\times 10^{-4}$.

In addition, taking the ratio of $B_{in} \times B_{out}$ for the two channels 
we measured, we obtain
$B(\chi_{c0}\rightarrow J/\psi\,\gamma)/B(\chi_{c0}\rightarrow \pi^0\pi^0)
=5.34\pm0.93\pm0.34$ 
(a number of minor systematic uncertainties cancel).
Using \cite{chi0_2pi0}, we then determine 
$B(\chi_{c0}\rightarrow J/\psi\,\gamma)
=(1.34\pm0.23_{stat}\pm0.09_{syst}\pm0.19_{PDG})\%$ and,
taking $\Gamma_{\chi_{c0}}$ from \cite{psigamma},  
$\Gamma_{\chi_{c0}\rightarrow J/\psi\,\gamma}=(131\pm26_{stat}\pm8_{syst}\pm18_{PDG})$~keV.
It is interesting to compare the E1 radiative transition of all three 
$\chi_{cJ}$ states.
The above measurement of $\Gamma_{\chi_{c0}\rightarrow J/\psi\,\gamma}$
is now in excellent agreement with the energy independent scaled quantities
$\Gamma_{\chi_{cJ}\rightarrow J/\psi\,\gamma}/q_{J}^3$ 
for $\chi_{c1}$ and $\chi_{c2}$ \cite{PDG}.


Summarizing, an interference pattern in the $\pi^0\pi^0$ cross section 
has been observed at the $\chi_{c0}$ mass.
An original analysis has been developed to extract the $\chi_{c0}$ resonance
parameters.
Combining the present study with our previous one of 
$\bar{p}p\rightarrow J/\psi\,\gamma$ \cite{psigamma}, important improvements in
the knowledge of the $\chi_{c0}$ are achieved.
The presented work proves that resonances can be observed and studied via
interference in final states dominated by non-resonant channels.
The developed analysis could be adopted in future studies,
such as $\pi^0\pi^0$ and $\pi^0\eta$ scans to search for possible 
$D\overline{D}$ bound systems at $\sim3700$~MeV.
Other two-body final states could be employed for the study of
charmonium singlet states via interference.

\begin{acknowledgments}
The authors thank the staff of their respective institutions and 
the Antiproton Source Department of the Fermilab Beams Division.
This research was supported by the US Department of Energy and 
the Italian Istituto Nazionale di Fisica Nucleare.
\end{acknowledgments}

%

\end{document}